\documentclass[a4paper]{article}

\usepackage[left = 4 cm, right = 4 cm]{geometry}
\usepackage[T1]{fontenc} 
\usepackage[utf8]{inputenc}
\usepackage{lmodern}
\usepackage{bm} 
\usepackage{amsmath, amsfonts, amssymb}
\usepackage{graphicx}
\usepackage{hyperref}
\usepackage{upgreek}
\usepackage[super, sort&compress]{natbib}
\usepackage{authblk}
\usepackage{todonotes, soul}
\usepackage{lineno}
\usepackage{setspace}

\newcommand{\jac}{Pt$_2$HgSe$_3$}

\author[1]{Konr\'{a}d Kandrai}
\author[1]{P\'{e}ter Vancs\'{o}}
\author[2]{Gerg\H{o} Kukucska}
\author[2]{J\'{a}nos Koltai}
\author[1]{Gy\"{o}rgy Baranka}
\author[1]{\'{A}kos Hoffmann}
\author[3]{\'{A}ron Pekker}
\author[3]{Katalin Kamar\'{a}s}
\author[1]{Zsolt E. Horv\'{a}th}
\author[4]{Anna Vymazalov\'{a}}
\author[1]{Levente Tapaszt\'{o}}
\author[1, $\dagger$]{P\'{e}ter Nemes-Incze}

\affil[1]{Centre for Energy Research, Institute of Technical Physics and Materials Science, 1121 Budapest, Hungary}
\affil[2]{ELTE E\"{o}tv\"{o}s Lor\'{a}nd University, Department of Biological Physics, 1117 Budapest, Hungary}
\affil[3]{Wigner Research Centre for Physics, Institute for Solid State Physics and Optics, 1121 Budapest, Hungary}
\affil[4]{Czech Geological Survey, 152 00 Prague, Czech Republic}
\affil[$\dagger$]{\small \emph{corresponding author, email: nemes@mfa.kfki.hu}}

\title{Signature of large-gap quantum spin Hall state in the layered mineral jacutingaite}

\begin{document}

\maketitle

\newpage

\begin{spacing}{1.5}

\section*{Abstract}
	\textbf{Quantum spin Hall (QSH) insulators host edge states, where the helical locking of spin and momentum suppresses backscattering of charge carriers, promising applications from low-power electronics to quantum computing.
	A major challenge for applications is the identification of large gap QSH materials, which would enable room temperature dissipationless transport in their edge states.
	Here we show that the layered mineral jacutingaite (Pt$_2$HgSe$_3$) is a candidate QSH material, realizing the long sought-after Kane-Mele insulator.
	Using scanning tunneling microscopy, we measure a band gap in excess of 100 meV and identify the hallmark edge states.
	By calculating the $\mathbb{Z}_2$ invariant, we confirm the topological nature of the gap.
	Jacutingaite is stable in air and we demonstrate exfoliation down to at least two layers and show that it can be integrated into heterostructures with other two-dimensional materials.
	This adds a topological insulator to the 2D quantum material library.}\\

	Keywords: Topological insulator, Low-dimensional materials, Quantum spin Hall effect (QSH), Scanning tunneling microscopy (STM)\\

	The QSH state \cite{Moore2010,Hasan2010} has first been realized experimentally, at cryogenic temperatures, in HgTe quantum wells \cite{Konig2007}.
	Interestingly, the prototype QSH insulator is actually graphene, when it was realized by Kane and Mele that its Dirac quasiparticles are gapped and characterized by a $\mathbb{Z}_2$ topological invariant if spin orbit coupling (SOC) is considered \cite{Kane2005a,Kane2005b}.
	However, the low SOC in graphene results in a gap of only a few $\upmu$eV, making its topological properties a mere theoretical curiosity.
	To realize a Kane-Mele insulator, a material is needed with the honeycomb lattice of graphene, but having large SOC \cite{Kane2005a}.
	In the last few years there has been a tremendous effort to find a layered material conforming to these requirements.
	From the point of view of applications, the candidate material forming this "heavy metal graphene", should ideally have the following characteristics.
	It should have a topological gap above room temperature, to enable room temperature dissipationless charge transport.
	The van der Waals bonding between the layers of the material should be weak enough \cite{Mounet2015} to enable exfoliation by the well known methods developed for 2D materials.
	This would enable integration into heterostructures with the vast numbers of other 2D quantum materials discovered to date \cite{Song2018c,Geim2013}.
	Such a combination with other 2D materials can enable a high degree of control over the edge states \cite{Song2018c}.
	For example, in proximity with 2D superconductors, Majorana quasiparticles could be formed \cite{Yan2018}.
	Lastly, it should be stable in air under ambient conditions, making the material widely usable.

	One possibility to realize a QSH system, is to increase the SOC in graphene by placing it in proximity to materials with a large atomic number \cite{Kou2013,Kou2014b,Alsharari2015}, either using adatoms \cite{Balakrishnan2013,Namba2018} or in a substrate \cite{Avsar2014,Wang2015b,Wang2016b}.
	The resulting SOC induced gap is of the order of 10 meV at best.
	An alternative is to find a material with an intrinsically large topological gap \cite{Ren2016}, such as a bismuth honeycomb layer on SiC \cite{Reis2016,Stuhler2019}, with a band gap of 0.8 eV.
	However, the crystal structure and therefore the topological properties of this bismuthene are linked to the SiC support, limiting its applicability.
	Similar constraints arise in the case of stanene \cite{Deng2018} and other group IV honeycomb layers and perhaps for bismuth (111) bilayers \cite{Drozdov2014,Peng2018}.

	Among materials that exist as freestanding single layers, the 1T' phase of transition metal dichalcogenides are predicted to be QSH insulators \cite{Qianh2014}.
	For MoS$_2$, WSe$_2$ and WTe$_2$ the hallmark edge states have been identified by scanning tunneling microscopy (STM) \cite{Xu2018b,Chen2018a,Ugeda2018,Tang2017c,Peng2017} and by charge transport measurements for WTe$_2$ \cite{Wu2017}.
	However, MoS$_2$ and WSe$_2$ are metastable and easily convert to the 2H phase \cite{Chen2018z}, while WTe$_2$ is stable in the 1T' polymorph, but rapidly oxidizes in air.
	None of the above examples are stable under ambient conditions, with the possible exception of Bi$_{14}$Rh$_3$I$_9$ \cite{Rasche2013a}.
	However, due to the complex crystal structure and ionic bonding between the layers \cite{Pauly2016}, it is not clear if it is possible to isolate a single layer of it.
	\\

	Here we present evidence via STM measurements that jacutingaite (\jac), a naturally occurring mineral \cite{Cabral2008,Vymazalova2012}, realizes a room temperature Kane-Mele insulator, satisfying all of the above criteria.
	By measuring on the basal plane of exfoliated multilayer crystals, we identify a bulk band gap and edge states within this gap, localized to monolayer step edges, showing a decay length of 5 \AA\ into the bulk.
	We reproduce the measured band gap and edge states by density functional theory calculations (DFT) of the monolayer.
	By calculating the $\mathbb{Z}_2$ invariant, we show that the band gap is expected to be topologically non-trivial, in accordance with the previous prediction of Marrazzo et al. \cite{Marrazzo2017}.
	Within our experiments \jac\ has proven to be stable under ambient conditions, on a timescale of months to a year, either as bulk or exfoliated crystals with a thickness down to 1.3 nm, equivalent to one or two layers.
	This is no surprise since jacutingaite is a mineral \cite{Cabral2008,Vymazalova2012}, therefore it should be stable not just under ambient but at pressures and temperatures relevant to geological processes.
	\\


	The sample we investigated was grown synthetically, as described previously \cite{Vymazalova2012}.
	For preparation and characterization details, see Supplementary section S1.
	Additionally, we have measured and calculated the Raman spectrum of bulk crystals, see Supplement S6.
	In the following, we focus on STM measurements of exfoliated thick crystals on a gold surface.
	The measurements were carried out in UHV at a base pressure of 5$\times$10$^{-11}$ Torr and a temperature of 9 K.

\begin{figure}[!htbp]
	\includegraphics[width = 1 \textwidth]{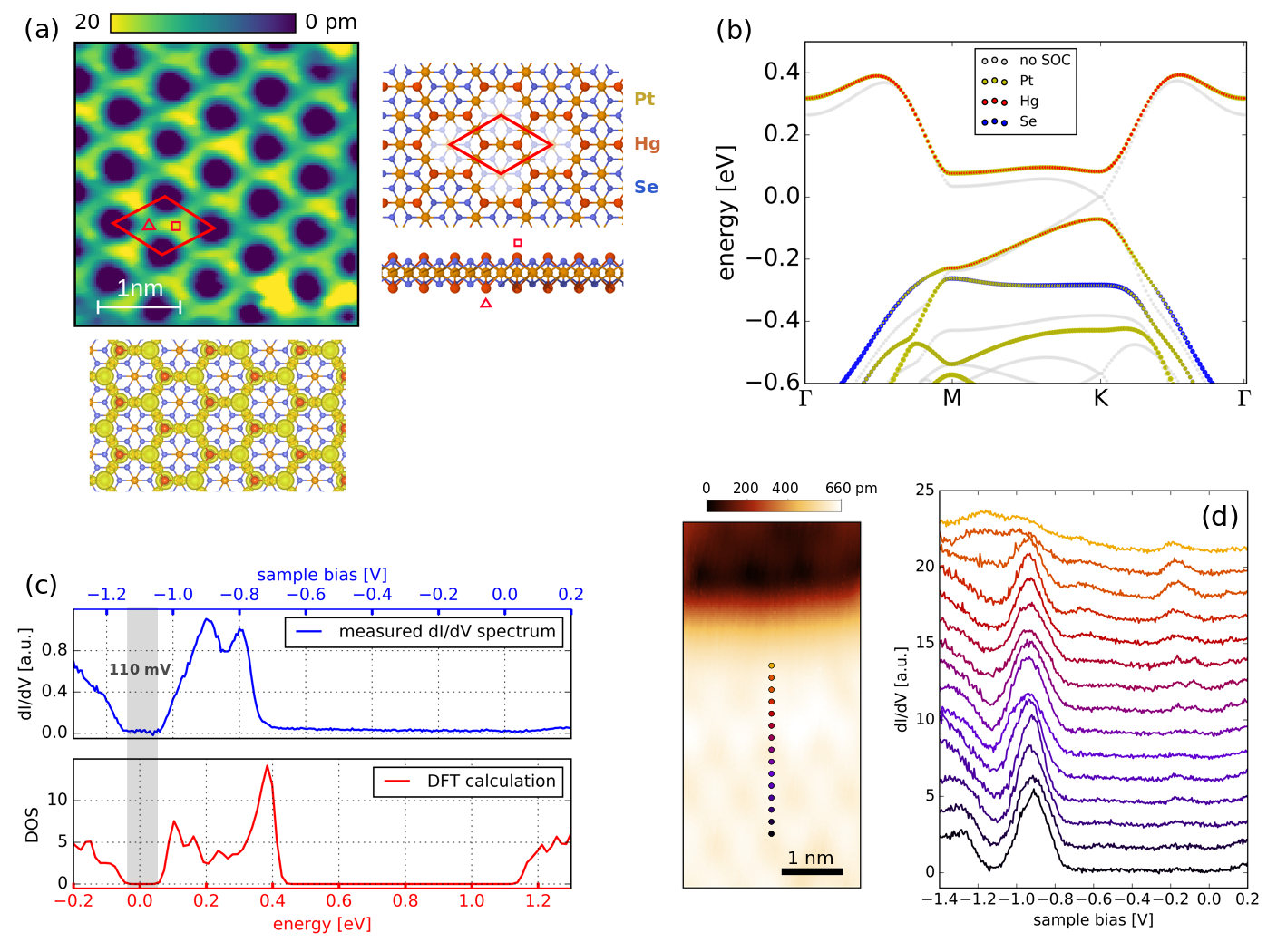}
	\caption{\textbf{Atomic and electronic structure of \jac.}
		\textbf{(a)} Atomic resolution, topographic STM image of \jac, stabilization parameters: 10 pA, -0.8 V.
		Sublattices are marked with a red triangle and rectangle, respectively.
		Right inset: atomic structure of \jac, top and side view.
		Bottom inset: Contour plot of the density of states within the conduction band in a 200 meV interval.
		\textbf{(b)} Band structure of \jac\ single layer, from DFT calculation, without (grey) and with (colored) SOC.
		Size and color of the dots is proportional to the weight of Pt, Hg or Se in the respective electronic state.
		\textbf{(c)} Comparison of measured $dI/dV(V)$ signal (blue) and calculated (red) density of states.
		The measurement was conducted on the defect free basal plane of \jac.
		The calculation is for a monolayer of \jac.
		Band gap highlighted in gray.
		\textbf{(d)} Measured $dI/dV$ spectra as a function of distance from a step edge on the basal plane.
		The spectra are offset for clarity.
		Topographic STM image of the step shown on the left side of the spectra.
		The positions of the spectra are shown by dots with the respective colors.
	}
\label{fig:atomic}
\end{figure}

	Jacutingaite is a ternary compound having a "sandwich-like" structure reminiscent of transition metal dichalcogenides, with a platinum layer between selenium and mercury.
	It can be regarded as "heavy metal graphene", since states around the SOC induced gap are localized on the honeycomb lattice formed by Pt and Hg atoms (see bottom inset in Fig. \ref{fig:atomic}a) \cite{Marrazzo2017}.
	Indeed, in the absence of SOC these bands give rise to a Dirac cone at the K points of the Brillouin zone (see Fig. \ref{fig:atomic}b).

	The atomic resolution STM images of the basal plane reflect this honeycomb structure, for an example see Fig. \ref{fig:atomic}a.
	The topographic image shows a sublattice symmetry broken graphene-like arrangement of the local density of states (LDOS), with the unit cell shown by a red rhombus.
	The unit cell size is measured to be 7.3 \AA, in agreement with the expected unit cell size (7.34 \AA) measured via X-ray diffraction \cite{Vymazalova2012}.
	Upon closer examination, we can observe a difference in the apparent height of the two sublattices, marked by red squares and triangles in Fig. \ref{fig:atomic}a.
	This sublattice symmetry breaking is a consequence of the buckled honeycomb nature of the Pt-Hg lattice.
	The buckling means that each inequivalent sublattice resides on opposing sides of the single layer, similarly to silicene or germanene.

	Measuring the differential tunneling conductivity ($dI/dV(V)$) on the defect free basal plane, reveals a bulk band gap of 110 mV, shown by the gray shading in Fig. \ref{fig:atomic}c.
	Importantly, if measured far away from any surface defects or edges, the $dI/dV$ signal goes to zero inside the gap, showing that this energy range is indeed devoid of electronic states.
	The measured LDOS is in excellent agreement with density functional theory (DFT) calculations of the monolayer, see red plot in Fig. \ref{fig:atomic}c.
	The 110 meV gap shown here is a best case scenario, where we purposely selected an area devoid of any surface defects.
	The large defect concentration of the basal plane (see Supplement S3) makes the local electronic structure inhomogeneous.
	In order to characterize the gap rigorously, we have measured the band gap from 982 individual spectra in an area 10x10 nm$^2$.
	The mean gap value was found to be 78 meV, with a standard deviation of 27 meV (for details see Supplement S4).
	The topological nature of the band gap is established by calculating the $\mathbb{Z}_2$ index (see Supplement S7).
	By comparing the red and blue plots in Fig. \ref{fig:atomic}c, we can immediately see that the calculated LDOS of the monolayer accurately reproduces the $dI/dV$ spectrum measured on the top layer of a bulk crystal.
	Also considering that the measurement is not reproduced by the calculated surface DOS of a 4 layer slab, suggests that the top \jac\ layer in our measurement is decoupled from the bulk (see supplement S7.1).
	This is supported by our STM measurements of the monolayer step height, which is found to be 0.7 \AA\ larger than the intrinsic interlayer distance of 5.3 \AA\ (see inset in Fig. \ref{fig:edge}a and Fig. 1c of the supplement).
	
	Although the measured LDOS is reproduced by the DOS of the monolayer, the sample is heavily $n$ doped.
	In case of the measured curve in Fig. \ref{fig:atomic}c, the Fermi level marked by zero bias is shifted above the conduction band, leaving the band gap at -1.15 eV.
	A possible source of the high $n$ doping might be defects or inhomogeneities in the bulk crystal (see Supplement S3).
	A strong indicator of these is the presence of PtSe$_2$ in the sample and that in the case of all crystals we observe a large number of adsorbates even on the freshly cleaved basal plane.
	Investigating the doping in exfoliated crystals down to the bilayer thickness, we find that the $n$ doping is considerably less, with the Fermi level being at least 0.5 V closer to the topological gap than for the bulk (see Supplementary figure 13 and 14).
	This points to inhomogeneities and defects as being the most likely cause of the doping, as well as the enlarged interlayer spacing.

	Having established the location of the band gap in the $dI/dV$ spectra, let us focus on investigating the presence of the predicted QSH edge states \cite{Marrazzo2017}.
	Other QSH material candidates, such as WTe$_2$ \cite{Peng2017}, Bi$_{14}$Rh$_3$I$_9$ \cite{Pauly2016}, and ZrTe$_5$ \cite{Wu2016} also reproduce the LDOS of the monolayer, when measuring the top of bulk crystals with STM.
	For these materials, monolayer steps on the bulk surface show the hallmark edge states residing in the band gap.
	In Fig. \ref{fig:atomic}d, we show individual $dI/dV$ spectra measured near a monolayer step edge on a thick flake, having hundreds of layers.
	The positions of the spectra are marked by similarly colored dots on the STM image of the step.
	At a position 2 nm away from the step edge, the spectra reproduce the LDOS measured deep in the bulk of the sample.
	Moving even closer to the edge, at a distance of $\sim$1 nm, the LDOS inside the band gap starts to increase, indicating the presence of an in gap state.
	An extra state localized to the edge also appears above the conduction band, at -0.2 V, which is a fingerprint of the edge structure.
	During our STM investigation, straight and atomically clean edges were always of the zigzag kind.
	Therefore, we checked the atomic and electronic structure of this edge orientation terminated by Se, Pt and Hg, by optimizing the atomic lattice of monolayer ribbons in DFT.
	The only atomic configuration that shows the hallmark edge state above the conduction band and is energetically stable, is a Se terminated zigzag edge (see Fig. \ref{fig:ribbon} and Supplementary section S8).
	Thus, we have used this trivial edge state above the conduction band to identify the type of zigzag edge present in the measurement.
	This allows us to accurately reproduce the LDOS of the edge in our calculations.

\begin{figure}[!htbp]
	\includegraphics[width = 1 \textwidth]{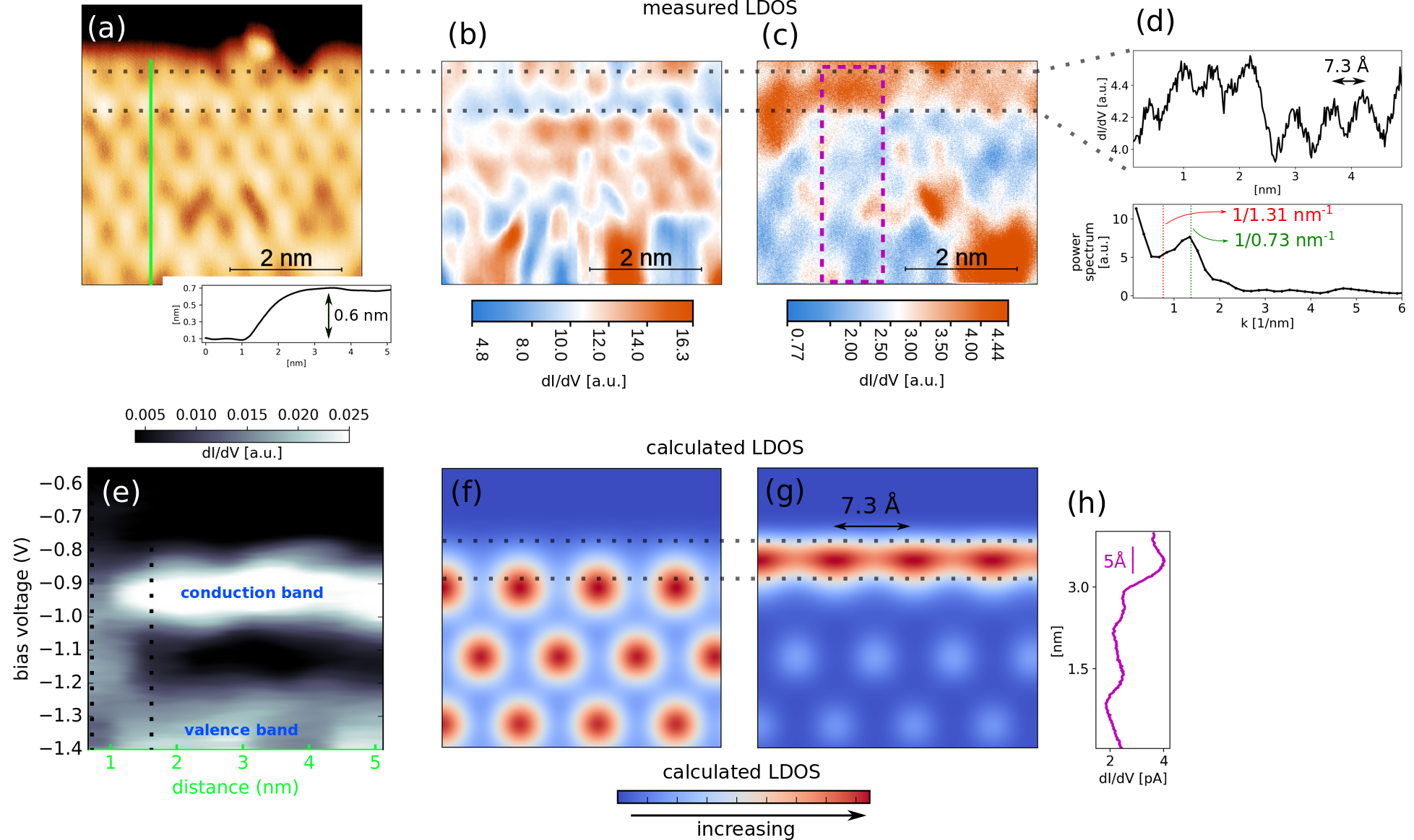}
	\caption{\textbf{Characterizing the edge state.}
		\textbf{(a)} Topographic STM image of a zigzag edge.
		Stabilization parameters: -0.85 V bias, 30 pA.
		$dI/dV$ spectra shown in (e) are measured along the green line.
		Black dotted lines mark the edge, as in (a-c, e-g).
		Inset: height section of the step.
		\textbf{(b)} $dI/dV$ image, measured in the same area as the topographic image in (a), outside the gap in the conduction band, at bias voltage -0.85 V.
		\textbf{(c)} $dI/dV$ image, measured in the same area as (a, b), at a bias voltage of -1.15 V inside the gap. The position of the edge state is marked between two dotted black lines.
		\textbf{(d)} Top: dI/dV signal modulation along the edge state.
		Section between the black dotted lines in (c).
		Bottom: Fast Fourier transform of the line section.
		\textbf{(e)} Plot of $dI/dV$ spectra measured as a function of distance from the edge. The spectra are recorded along the green line in (c).
		\textbf{(f)} Calculated LDOS of the conduction band.
		\textbf{(g)} Calculated LDOS of the edge state, using a broadening of 2.6 \AA.
		LDOS periodicity along the edge is equal to the unit cell size (shown by arrowed black line).
		Edge state LDOS is concentrated between the dotted black lines.
		\textbf{(h)} Averaged section across the edge state within the purple dotted box shown in (c).
		The decay of the edge state into the bulk is of the order of 5 \AA, the same as the decay in the calculation: (g)
		For extended data, see section 10 of the supplement.
	}
\label{fig:edge}
\end{figure}

	In the following we examine in more detail the increased LDOS near the monolayer step edge.
	In Fig. \ref{fig:edge}c we show an image of the $dI/dV$ signal at a voltage inside the gap, measured along an edge shown in Fig. \ref{fig:edge}a.
	An increased $dI/dV$ signal indicates an increased LDOS near the step.
	In all panels on Fig. \ref{fig:edge} the black dotted lines mark the position of the edge.
	The decay of the edge state into the bulk is found to be of the order of 5 \AA, in agreement with prediction \cite{Marrazzo2017}.
	Taking a section between the dotted black lines (Fig. \ref{fig:edge}d), one observes that the edge LDOS is modulated by the atomic periodicity, as expected for a topological edge state \cite{Pauly2015,Stuhler2019}.
	A further hallmark of topological edge states is that the state is not perturbed by the presence of a defect, visible in the top-right area of Fig. \ref{fig:edge}a.
	If backscattering would take place due to the defect this would result in a modulation of the local density of states along the edge.
	The wavelength of this modulation is determined by the change is crystal momentum of the scattered electron, which can be obtained from the dispersion relation along the edge, shown in Fig. \ref{fig:ribbon}a.
	The voltage used in the measurement (-1.15 V) corresponds to an energy in the middle of the gap.
	At this energy, the change in crystal momentum would result in a periodicity of 13.1 \AA\ related to intra-band scattering \cite{Stuhler2019}.
	To check the presence of backscattering, we show the Fourier transform of the dI/dV signal along the edge in Fig. \ref{fig:edge}d.
	We observe the peak corresponding to 1/0.73 nm$^{-1}$ unit cell periodicity, but the peak for backscattering is clearly absent.
	This conclusion is further strengthened by additional Fourier analysis on a longer, irregular edge (see supplement S5).
	This analysis is essentially a 1D analog of probing the suppression of backscattering on the 2D surface state of strong topological insulators by STM measurement of quasiparticle interference patterns \cite{Roushan2009}.
	
	Finally, comparing the $dI/dV$ images with the calculated LDOS map inside the topological gap (Fig. \ref{fig:edge}g) and of the complete valence band (Fig. \ref{fig:edge}f), we find that there is good agreement with the measurements.
	The calculated LDOS maps reproduce both the atomic periodicity along the edge state, as well as its decay length of 5 \AA.
	With such a small decay length, it is expected that the edge state would start to develop at defect sites inside the basal plane, such as in the bottom-right corner of Fig. \ref{fig:edge}c.
	A better example of this effect can be observed in the supplementary Fig. 4d.

\begin{figure}[!htbp]
	\centering
	\includegraphics[width = 0.6 \textwidth]{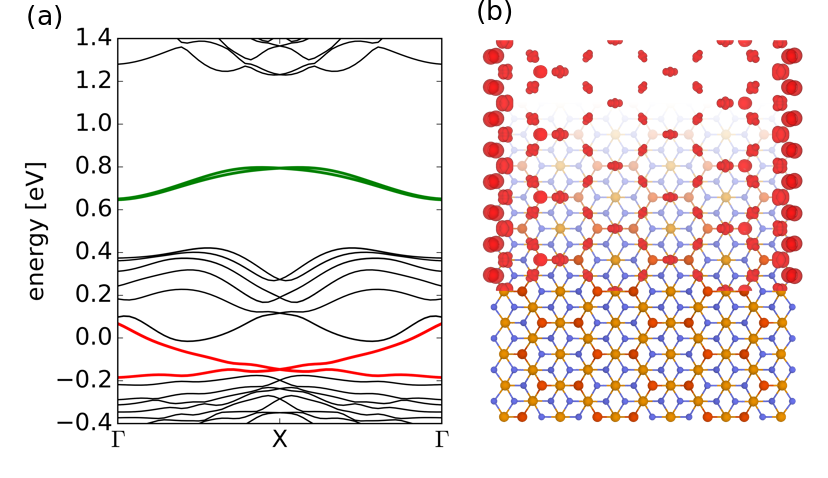}
	\caption{\textbf{\jac\ nanoribbon. Topological edge states.}
		\textbf{(a)} Band structure of a 3.2 nm wide zigzag ribbon, calculated using DFT.
		Topological edge state is shown in red, while the trivial edge state above the conduction band is shown in green.
		\textbf{(b)} LDOS contour plot of the topological edge state integrated over the whole topological band.
	}
\label{fig:ribbon}
\end{figure}

\begin{figure}[!htbp]
	\centering
	\includegraphics[width = 0.7 \textwidth]{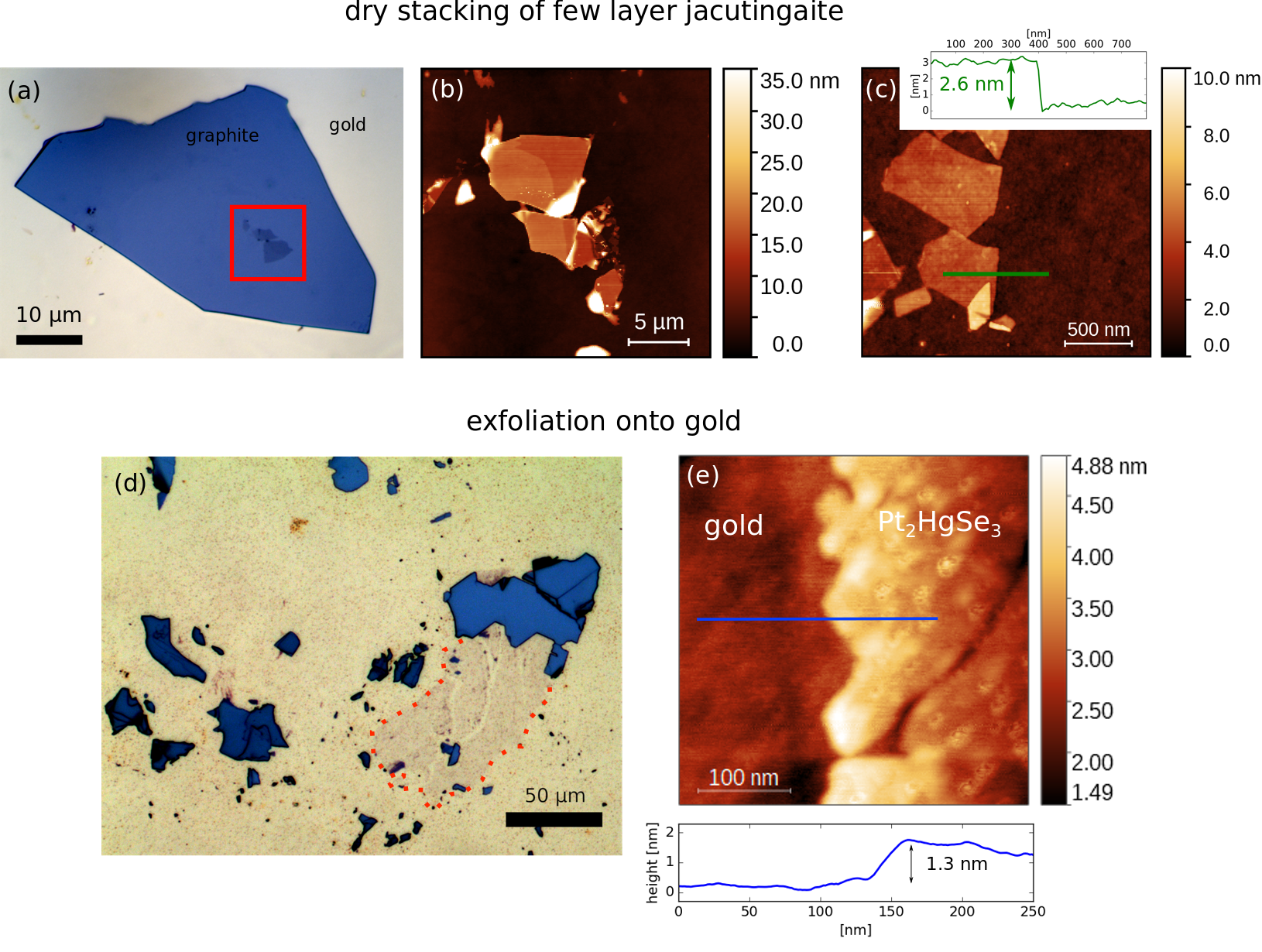}
	\caption{\textbf{Exfoliation of \jac.}
		\textbf{(a)} Stack of jacutingaite on graphite, prepared by dry stacking.
		\textbf{(b)} AFM image of the flake before the transfer supported on a PMMA substrate.
		\textbf{(c)} AFM image of a thin flake, having a thickness of 2.6 nm, corresponding to approximately 5 single layers. Inset: height section of the flake along the green line.
		\textbf{(d)} Exfoliation of jacutingaite onto a gold (111) surface.
		The thinnest flakes are marked by the red dotted line.
		\textbf{(e)} AFM image of the thinnest flakes, inside the area marked with red in (d).
		Inset: height section along the blue line.
	}
\label{fig:exfoliation}
\end{figure}

	The relatively weak van der Waals bond between the monolayers of \jac\ makes it possible to exfoliate the material, potentially to the monolayer limit \cite{Mounet2015}.
	We demonstrated this possibility by using the standard "scotch tape method" to exfoliate thin flakes onto a SiO$_2$ substrate or a polymer stack, as used in dry stacking of 2D materials \cite{Pizzocchero2016} (see Fig. \ref{fig:exfoliation}a-c).
	Using dry stacking, it should be possible to place \jac\ on the surface of a high T$_\mathrm{c}$ superconductor, enabling the investigation of high temperature Majorana zero modes \cite{Yan2018}.
	The thinnest crystals we were able to prepare by conventional scotch tape exfoliation onto SiO$_2$ substrates was 5 layers.
	However, these crystals have lateral sizes below 1 $\upmu$m (see Fig. \ref{fig:exfoliation}c), severely limiting their usefulness.
	Exfoliating onto fresh gold surfaces \cite{Magda2015} increases the lateral size of the flakes significantly and their thickness, measured by AFM is 1.3 nm (see Fig. \ref{fig:exfoliation}d, e).
	However, these thin flakes are found to be highly disordered.
	For more details see supplementary section S9.
	These results show that it should be possible to exfoliate single layers of \jac\ onto SiO$_2$ and especially gold substrates, but the material homogeneity and defect density of the bulk crystals needs to be improved significantly.
	Further improvements in crystal quality could also be a key to probing the dual topological nature \cite{Facio2019,Marrazzo2019} of \jac\, such as in the case of Bi$_2$TeI \cite{Avraham2017}.
	This is because \jac\ is predicted to not only be the long sought after Kane-Mele insulator, but in it's bulk form it is also a topological crystalline insulator and a $\mathbb{Z}_2$ insulator \cite{Marrazzo2019,Cucchi2019,Facio2019}.
	\\

	One of the most promising QSH materials is monolayer 1T'-WTe$_2$, but the  chemical stability of \jac\ in air and it's band gap above room temperature, clearly sets it aside.
	The main difference being that WTe$_2$ rapidly oxidizes under ambient conditions and shows the QSH effect only below a temperature of 100 K \cite{Wu2017}.
	Our results establish that jacutingaite is a new and widely accessible platform to explore the properties of helical one dimensional electron systems \cite{Novelli2018,Stuhler2019} and should be available for charge transport measurements, even in the monolayer, if the defect concentration and sample homogeneity can be improved.
	Recent theoretical studies highlight the possibility of superconductivity in doped \jac\ \cite{Wu2018d}, this could open a way to explore the coexistence of topological edge states in proximity to a superconductor in the same material system.
	Additionally, a non zero $\mathbb{Z}_4$ index \cite{Vergniory2019} makes \jac\ a fertile playground to explore higher order topology.
	In our samples the Fermi level is already shifted above the type-II van Hove singularity where superconductivity is expected, possibly due to the presence of lattice defects.
	Our results hint at the possibility that tuning the composition, may be an effective tool to control the doping of \jac, similarly to quaternary topological insulators \cite{Arakane2012}.

\subsection*{Data availability}

	The datasets generated during and/or analysed during the current study are available from the corresponding author on reasonable request.

\subsection*{Acknowledgments}

	L.T. acknowledges financial support from the ERC Starting grant NanoFab2D.
	P.N.I. acknowledges support form the Hungarian Academy of Sciences, Lend\"{u}let Program, grant no: LP2017-9/2017.
	The work was conducted within the Graphene Flagship, H2020 Graphene Core2 project no. 785219 and the Quantum Technology National Excellence Program (Project No. 2017-1.2.1-NKP-2017-00001).
	Work was supported by the National Research, Development and Innovation Office (Hungary) grant No. FK 125063 (\'{A}.P., Ka.K.), K-115608 (J.K. and G.K.), KH130413 (V.P.) and K108753 (L.T.).
	J.K. and G.K acknowledge the ELTE Excellence Program (1783-3/2018/FEKUTSTRAT) supported by the Hungarian Ministry of Human Capacities.
	V.P. acknowledges the Janos Bolyai Research Scholarship of the Hungarian Academy of Sciences.
	J.K. was supported by the UNKP-19-4 New National Excellence Program of the Ministry for Innovation and Technology.
	A.V. acknowledges financial support from the Grant Agency of the Czech Republic (project No. 18-15390S).
	We acknowledge NIIF for awarding us access to computing resources based in Hungary at Debrecen.
	Ka.K. acknowledges grant no. VEKOP-2.3.2-16-2016-00011.

\subsection*{Associated Content}
	The Supporting Information is available free of charge on the ACS Publications website at DOI: ...
	
	Supplementary sections 1 to 9:
	Detailing the sample preparation and STM measurement, additional information on defects, band gap statistics, measurements on irregular, monolayer edges, Raman measurements, density functional theory calculation details, various edge configurations, details on exfoliation.

\subsection*{Author contributions}
	
	Ko.K. did the exfoliation experiments and STM measurements, with the supervision of P.N-I.
	\'{A}.H. helped with sample preparation.
	A.V. provided the sample.
	P.V., G.K. and J.K. performed the DFT calculations.
	G.B., \'{A}.P. and Ka.K. performed the Raman measurements, while G.K. calculated the Raman spectrum, under the supervision of J.K.
	A.V. and Z.E.H. performed the XRD measurement.
	P.N-I. conceived the project and coordinated it together with L.T.
	P.N-I. wrote the manuscript, with contributions from all authors.

\newpage


\end{spacing}

\end{document}